\documentclass[twocolumn,preprintnumbers,superscriptaddress,amsmath,amssymb,aps]{revtex4-1}
\usepackage{graphicx,caption,subcaption,float}
\begin{document}

\title{Ergodicity Breaking and Localization}

%
%

 \author{Elvis Geneston}
 \affiliation{Department of Physics, La Sierra University, Riverside, CA, USA}
        
 \author{Rohisha Tuladhar}
 \email{raisha.t@gmail.com}
 \affiliation{Center for Nonlinear Science, University of North Texas,
 	P.O. Box 311427, Denton, Texas 76203-1427}
 
  \author{M.T. Beig}
  \affiliation{Center for Nonlinear Science, University of North Texas,
  	P.O. Box 311427, Denton, Texas 76203-1427}
 
 \author{Mauro Bologna}
 \affiliation{Instituto de Alta Investigaci\'{o}n, Universidad de Tarapac\'{a}, Casilla 6-D, Arica, Chile}

 \author{Paolo Grigolini}
 \email{Paolo.Grigolini@unt.edu}
 \affiliation{Center for Nonlinear Science, University of North Texas,
 	P.O. Box 311427, Denton, Texas 76203-1427}

\begin{abstract}
We study the joint action of the non-Poisson renewal events (NPR) yielding Continuous Time Random Walk (CTRW) with index $\alpha < 1$ and two different generators of Hurst coefficient $H \neq 0.5$, one generating 
fractional Brownian motion  (FBM) and another  scaled Brownian motion (SBM). We discuss the ergodicity breaking emerging from these joint actions and we find
that in both cases the adoption of time averages leads to localization. In the case of the joint action of NPR and SBM, localization occurs when SBM would produce sub-diffusion. The joint action of NPR and FBM, on the contrary, may lead to localization when FBM is a source of super-diffusion. The joint action of NPR and FBM is equivalent to extending the CTRW to the case where the jumps of the runner are correlated and we argue that the the memory-induced localization requires a refinement of the theoretical perspective about determinism and randomness. 

\begin{description}
	
	\item[PACS numbers]
	\verb+ 	87.10.Mn, 02.50.-r, 05.40.Fb, 05.40.-a+ 
	
\end{description}
\end{abstract}

\pacs{Valid PACS appear here} 
\maketitle

\section{Introduction}  \label{introduction}

The proper evaluation of averages in statistical physics
requires the recourse to the  Gibbs idealization of infinitely many copies of the same system, namely, to the Gibbs ensemble that is of fundamental importance for the theoretical predictions.
In practice, when only the observation of a single system is possible, this ideal ensemble average is assumed to be identical to the more accessible average in time on the same system. This is the ergodic assumption \cite{lebowitz} that 
the tracking of single molecules proves to be frequently violated \cite{review}, a phenomenon called \emph{ergodicity breaking}.  Establishing a good theoretical command of ergodicity breaking is of fundamental importance when we move from physics  to biology, to physiology and to sociology, since ergodicity breaking seems to be a general property of these complex systems.   

\subsection{Continuous Time Random Walk}

It has to be stressed, however, that one of the origins of the ergodicity breaking observed in molecular diffusion in biological cells is the occurrence of non-Poisson renewal (NPR) events, which generate non-stationary correlation functions, characterized by aging \cite{review} in a form that is not to be confused with the non-stationarity of the slow approach to thermodynamical equilibrium. 

This  form of ergodicity breaking was very clearly illustrated in \cite{sokolov1,reference}. The diffusion process of a molecule that as a result of ensemble average would yield 
 $\left<x^{2}(t)\right> \propto t^{\alpha}$ 
with $\alpha < 1$, namely the sub diffusional scaling $\eta = \alpha/2$,  was studied by the authors of Refs. \cite{sokolov1,reference} making averages in time rather than on infinitely many copies of the same system. These authors showed that the second moment evaluated in time does not yield the scaling $\alpha/2$.  They got  instead 
the result \cite{note}

\begin{equation} \label{definition}
\left<\overline{\delta^2}(\Delta) \right> = \frac{1}{L-\Delta} \int_{0}^{L-\Delta}  \left<\left[x(t) - x(t+\Delta)\right]^2 \right> dt \propto \frac{\Delta}{L^{1-\alpha}},
\end{equation}
where $L$ is the length of the time series analyzed.
This impressive result indicates that the adoption of time average
turns a sub-diffusion into a normal diffusion process with the 
surprising property, however, that the intensity of the second moment becomes weaker and weaker upon increasing the length of the time series. 
The same kind of dependence on the length of the time series holds true for the power spectrum $P(f)$ of the renewal generators of sub-diffusion,
very well described \cite{margolin,mirko} by
\begin{equation} \label{true1/f}
P(f) \propto \frac{1}{f^{2-\alpha}} \frac{1}{L^{1-\alpha}}.
\end{equation}
Notice that we use the condition $\alpha < 1$ throughout the whole paper and that we adopt the notation $\left<O\right>$ and $\overline O$ to denote ensemble and time average, respectively. We shall adopt the notation 
$\left<\overline O\right>$ to denote ensemble averages of time averages. 

The result of Eq. (\ref{definition}) found in Refs. \cite{sokolov1,reference} is based on the  model of anomalous diffusion, called Continuous Time Random Walk (CTRW) \cite{ctrw}.  The runner makes a sequence of jumps $\xi(1), \xi(2),  ....\xi(n), ...$. These are uncorrelated Gaussian fluctuations, with vanishing mean value $\left<\xi\right>=0$ and  a finite second moment $\left<\xi^2\right>$. The number of events $n$ is interpreted \cite{sokolov} as operational time. In the clock time the time interval between two operational events is described  by the distribution density
\begin{equation}
\lim_{\tau \rightarrow \infty} \psi(\tau) \propto \frac{1}{\tau^{1 + \alpha}},
\end{equation}
with $0 < \alpha < 1$. 
The authors of Ref. \cite{operationaltime} derived this distribution density for neurophysiological events assuming a logarithmic relation between the clock time
and the subjective or psychological time of the individuals. To make our picture as general as possible, 
including other possible sources of complexity, we use
the term operational time to denote the number of renewal events, rather than psychological time as it would be natural in the case of neurophysiological processes \cite{operationaltime}.  We shall focus on the condition $ n  \gg 1$
and for this reason we shall interpret $n$ as being the continuous time, hereby denoted by the symbol $\tau_{\psi}$.  The clock time is continuous, but the numerical calculations convert it into a discrete time that can be interpreted again as continuous in the long-time limit. The theoretical interpretation of the clock-time region very close to the origin $t= 0$  forces us to consider again the discrete nature of operational time. This will be done hereby by introducing the parameter $L_0$, of the order of the numerical time step $\Delta t$, below which no event occurs.

\subsection{Extension of Continuous Time Random Walk} \label{extension}

We extend CTRW to the condition where 
in the operational time scale the fluctuation perceived by the runner is not random but correlated, and the decay of the correlation function of the fluctuation $\xi$ is extremely slow.  Let us consider
the stationary and normalized correlation function
\begin{equation} \label{correlation1}
\Phi_{\xi}(|t_1-t_2|) = \frac{\left<\xi(t_1)\xi(t_2)\right>}{\left<\xi^2\right>},
\end{equation}
with $\left<\xi\right >= 0$. 
We assign to the correlation function the analytical form
\begin{equation} \label{correlation2}
\Phi_{\xi}(\tau \equiv |t_1- t_2|) = \left(\frac{T_B}{\tau + T_B}\right)^{\nu}.\end{equation}
The correlation time $\tau_c$ of the correlation
function is established by evaluating  the Laplace 
transform of $\Phi_{\xi}(\tau)$,
\begin{equation}
\hat{\Phi}_{\xi}(u) \equiv \int_{0}^{\infty} \exp \left(-ut \right)
\Phi_{\xi}(\tau),
\end{equation}
which, after setting $u=0$, yields
 
 \begin{equation}
 \tau_c = \begin{cases}
 \frac{T_B}{\nu -1} & \nu>1 \\
 \infty & 0< \nu <1. \\
 \end{cases}
 \end{equation}
 We  focus on the case of the very slow approach to equilibrium with $0 < \nu < 1$. This correlation function is not integrable, thereby making the correlation time infinite.
This generalized form of CTRW, with the random fluctuation $\xi$ of CTRW replaced by a correlated fluctuation,  is equivalent to making the random walker run  in the operational time regime  the well known fractional Brownian motion (FBM), as discussed in Refs. \cite{cakir,arkady}, rather than the ordinary  Brownian motion.    

It is important to stress that FBM is ergodic in the sense that both the ensemble average and the time average yields $\left<x^{2}(t)\right> \propto t^{2H}$ and $\left<\overline{\delta^2}(\Delta) \right> \propto \Delta^{2H}$, with $0 < H < 1$.   However, the authors of Ref. \cite{deng} found that this ergodic regime is realized after a very extended non-ergodic regime that becomes perennial in the limit $H \rightarrow 1$. It is possible that the main result of this article, see Section \ref{mainresult} may have a connection with this extremely slow convergence to the ergodic regime. 

To make it possible for the readers to appreciate the role of FBM memory we find  it convenient to discuss also the joint effect of NPR events and another form of ergodicity-breaking process  called Scaled Brownian Motion
(SBM) \cite{sum}. This form of anomalous diffusion is attracting an increasing  interest  by the researchers working on anomalous diffusion in biological cells \cite{review,scaledfbm,safdari,bodrova} and granular gases \cite{granular}. 
The SBM \cite{scaledfbm}
generates the fluctuation $\xi(t)$ according to the prescription
\begin{equation} \label{prescriptionofsokolov}
\xi(t) = \sqrt{k t^{2H-1}} \xi_R,
\end{equation}
where $k$ is a constant with the proper dimension  and $\xi_R$ a random Gaussian noise, thereby yielding the time dependent diffusion coefficient $D(t)$ 
given by 
\begin{equation} \label{timedependentdiffusioncoefficient}
D(t) = k t^{2H-1}
\end{equation}
and leading the second moment of the diffusing variable $x(t)$, defined by $\dot x = \xi$,  to increase in time as
\begin{equation} \label{definingSBM}
\left<x(t)^2\right> \propto \int_{0}^{t} dt' D(t') \propto t^{2H}. 
\end{equation}

\subsection{Main result of this paper} \label{mainresult}

The main result of this paper is the theoretical and numerical discovery that  the joint action of FBM memory and NPR events yields the localization of single trajectories. When the correlation function  of Eq. (\ref{correlation2}) is sufficiently slow the fluctuations of single trajectories generated by the extended form of CTRW described in Section \ref{extension} tend to vanish, thereby making the position $x(t)$ virtually time independent in the time asymptotic limit. This is equivalent, to some extent that is explained in text, to replacing the factor $1/L^{1-\alpha}$  in  Eq. (\ref{definition}) with $1/L^{\gamma}$ where $\gamma \equiv 2H(1-\alpha) $,  with the effect of 
generating a form of ergodicity breaking similar to that of  Eq. (\ref{definition}), in the sense that the intensity of  $\overline{\delta^2}(\Delta)$ is a decreasing function of the total length $L$ of the time series analyzed.  The parameter $\gamma$ is not confined to $\gamma < 1$ and  for $\alpha < 0.5$ a critical value of $\nu$, $\nu_c$,  exists yielding $\gamma > 1$ for $\nu < \nu_c$
and making the intensity of  $\overline{\delta^2}(\Delta)$ vanish. We refer to this phenomenon as \emph{memory-induced localization}.  It is important to stress that this form of localization, being a property of single trajectories,
is compatible with the spreading of different trajectories generating the non vanishing scaling $\eta = \alpha H$.

Using the same theoretical approach, this article sheds light also on the formal result of the negative scaling coefficient $\eta$ found in the earlier work of 
Ref. \cite{bologna}. In this paper we  see that this earlier result is explained by the joint action of NPR and SBM,  with the effect of making $\overline{\delta^2}(\Delta)$ vanish.  However, this form of joint action is less surprising than the main result of this article, because the joint action of SBM fluctuations and NPR events yields localization when both processes yield sub-diffusion.

The outline of this paper is as follows. In Section \ref{renewal} we review the connection between NPR processes and the out of equilibrium cascade of events produced by the preparation of these systems, with an event occurring at $t = 0$. Section \ref{clock} is devoted to the discussion of the joint action of NPR events and FBM fluctuations.  In Section \ref{secondform} we discuss the joint action of NPR events and SBM fluctuations. Finally we devote 
Section \ref{concludingremarks} to concluding remarks.

\section{Renewal} \label{renewal}

To generate the cascade of renewal events in this paper we adopt an algorithm based on the idealized Manneville  map \cite{compression} prescription. In this case the survival probability, namely, the probability that a new event occurs at a time interval larger than $t$ from an earlier event is given by the function $\Psi(t)$ defined by
\begin{equation}
\Psi(t) = \left(\frac{T}{T + t}\right)^{\alpha},
\end{equation}
where the parameter $T$ defines the short-time scale, not showing yet the asymptotic complexity. 
The survival probability $\Psi(t)$ fits the normalization condition
\begin{equation}
\Psi(0) = 1,
\end{equation}
which implies that an event certainly occurs at a given $t> 0$. 
The corresponding waiting time distribution density is given by
\begin{equation} \label{manneville}
\psi(t) =  \frac{\alpha T^{\alpha}}{(t+T)^{1+\alpha}},
\end{equation}
which in the asymptotic time limit clearly shows its inverse power law complexity with the analytical form
\begin{equation} \label{manneville1}
\psi(t) = \frac{\Lambda^{\alpha}}{t^{1+\alpha}},
\end{equation} 
where
\begin{equation} \label{lambda}
\Lambda \equiv {\alpha}^{1/\alpha} T.
\end{equation}

Let us consider the case of infinitely many realizations with the condition that all of them have an event at time $t = 0$. The rate of events at time $t$ is given by
\begin{equation} \label{eli}
R(t) = \sum_{n=1}^{\infty} \psi_{n}(t),
\end{equation}
where $\psi_{n}(t)$ is the probability density that an event occurring  at time $t$ is  the last of a series of $n-1$ earlier events. The renewal condition leads to the mathematical definition of $\psi_{n}(t)$ through the iterative relation
\begin{equation} \label{timeconvolution}
\psi_{n}(t) = \int_{0}^{t} \psi_{n-1}(t') \psi_{1}(t-t'), 
\end{equation}
where $\psi_{1}(t) = \psi(t)$ with $\psi(t)$ given by Eq. (\ref{manneville}). The time convolution of Eq. (\ref{timeconvolution})  makes it possible to derive the Laplace transform of the right hand side of Eq. (\ref{eli}) and so of $R(t)$. By inverse-Laplace transforming this result, see e.g. Ref. \cite{reference}, we are led to 
\begin{equation}  \label{feller}
R(t) = \frac{A}{t^{1-\alpha}},
\end{equation} 
with
\begin{equation} \label{valueofA}
A \equiv \frac{\alpha}{\Lambda^{\alpha} \Gamma(\alpha) \Gamma(1-\alpha)}.
\end{equation}
The expression of Eq. (\ref{feller}) is the well known cascade of Feller events \cite{fellerbook}, which is the clear sign of the of the fact that there is no characteristic time scale for the dynamics of the system, when 
$\alpha < 1$.  

It is important to notice that, as earlier mentioned,  $n$ can be interpreted as a discrete time, usually called operational time.  In the asymptotic limit the operational time  $n \rightarrow \infty$ becomes identical to a continuous time that we denote as
$\tau_{\psi}$.  The numerical simulation of this article is based on the adoption of the finite time step $\Delta t = 1$. The evaluation of  $\overline{\delta^2}(\Delta)$ according to the prescription of Eq. (\ref{definition})
is done with moving windows of size $\Delta$. In Eq. (\ref{definition}) the left side of the window closest to the preparation  of system is located at $t=0$. To make easier for the reader to understand the phenomenon of memory-induced localization we modify this definition adopting as bottom limit of integration $L_0 > 0$ rather than $t= 0$. This definition rests on the interpretation of discrete operational time as counting the occurrence of events and on the fact that in addition to the preparation event no event can occur at time shorter than $L_0$. The order of magnitude of $L_0$ is, of course, $\Delta t =1$.

 \section{Joint action of NPR events and FBM fluctuation} \label{clock}
 
 The first form of joint action of two different sources of anomalous 
diffusion  discussed in this article is an extension of CTRW.  In the ordinary CTRW when the runner jumps, she makes jumps in the positive or negative direction according to the sign of the fluctuation $\xi$ and 
this fluctuation is random. In the extended CTRW of this article the runner sees in her operational time a correlated fluctuation $\xi$, as a consequence  of the infinite memory of the FBM generating fluctuations.  In other words, we assume that the runner in her operational time $\tau_{\psi}$ makes a diffusion described by FBM.  
In the clock time, in the extended time region between two events, the runner does not move.  This has the effect of producing  the time scale dilatation 
\begin{equation} \label{truescaling}
t = \tau_{\psi}^{1/{\alpha}},
\end{equation}
thereby favoring sub-diffusion. This problem has been already studied in the earlier work of Refs. \cite{sokolov2,sokolov3,sokolov4}. 

However, we focus on the case $H>0.5$ which in the operational time scale yields super-diffusion. For this reason we think that the localization 
effect revealed by the theoretical and numerical analysis of this article is a surprising property. 

\subsection{Review of the dynamical origin of FBM}

To make this paper as self contained as possible, let us review the derivation of FBM done in the earlier work of Refs. \cite{cakir,arkady}. We move from the equation  of motion
\begin{equation} \label{tointegrate}
\dot x \equiv \frac{dx}{dt} = \xi(t).
\end{equation}
In this paper to generate the trajectory $x(t)$ yielding the fluctuation $\xi(t)$ we use the algorithm of Ref. \cite{misraelvis}. This not only allows us to establish an approach
equivalent to the dynamical approach to FBM of the earlier work of Refs. \cite{cakir,arkady} but it also makes it possible for us to make a comparison with SBM of Eq. (\ref{prescriptionofsokolov}). After recording $\xi(t)$ 
we evaluate its correlation function using the time average and, resting on the ergodicity property,  we establish numerically the parameter $T_B$  of Eq. (\ref{correlation1}) and the mean quadratic value $\left<\xi^2\right>$. 

Notice that although the prescriptions of this subsection will be applied to the operational time $\tau_{\psi}$, for notational simplicity we adopt the conventional symbol $t$ for time.  The fluctuation $\xi(t)$ is characterized by the
stationary correlation function of Eq. (\ref{correlation1}) 
and Eq. (\ref{correlation2}). We integrate Eq. (\ref{tointegrate}), square the result, make the ensemble average and we use the property that the correlation function depends on  $|t_1-t_2|$ (see Eq. (\ref{correlation1})) to get
\begin{equation}
\left<x^2(t)\right> = 2 \left<\xi^2\right> \int_{0}^{t} dt' \int_{0}^{t'} dt'' \Phi_{\xi}(t''). 
\end{equation}
Thus we obtain that for $t \rightarrow \infty$
\begin{equation}
\left<x^2(t)\right>  = c t^{2H},
\end{equation}
where
\begin{equation} \label{definitionofdelta}
H \equiv 1 - \frac{\nu}{2}
\end{equation}
and
\begin{equation} \label{definitionofc}
c \equiv \begin{cases}
 \frac{2\left<\xi^2\right> T_B^{\nu}}{(1-\nu)(2-\nu)} & H\neq 0.5 \\
 \left<\xi^2\right> & H= 0.5. \\
 \end{cases}
\end{equation}

\subsection{Time averages} \label{timeaverage}
The evaluation of the second moment through time averages is given by 
\begin{equation} \label{ideal}
\overline{\delta^2}(\Delta) = \frac{1}{L-\Delta} \int_{L_0}^{L-\Delta} \left[x(t) - x(t+\Delta)\right]^2 dt,
\end{equation}
where $L$ is the total length of the time series of the diffusional variable $x(t)$. Note that, as earlier stated, to take into account the discrete nature of the operational time the bottom limit of time integration is given by $L_0$, which is of the order of $1$.

We adopt the approximation of keeping the length $\Delta$ of the mobile window much smaller than the total length $L$, thereby replacing Eq. (\ref{ideal}) with\begin{equation} \label{real}
\overline{\delta^2}(\Delta) = \frac{1}{L} \int_{L_0}^{L} \left[x(t) - x(t+\Delta)\right]^2 dt.
\end{equation}

After evaluating the time average over a single trajectory we 
make the time average over another trajectory, and so on.
Then we evaluate an ensemble average over the time averages 
and we rewrite Eq. (\ref{real}) as
\begin{equation} \label{ensemble}
\left<\overline{\delta^2}(\Delta)\right> = \frac{1}{L} \int_{L_0}^{L} \left<\left[x(t) - x(t+\Delta)\right]^2\right> dt.
\end{equation}

Note that $x(t)$ changes only when an event occurs in the operational time. As a consequence
\begin{equation} \label{dependenceonH}
\left<\left[x(t) - x(t+\Delta)\right]^2\right> = c n(t+\Delta,t)^{2H},
\end{equation}
where $c$ is given by Eq. (\ref{definitionofc}) and $n(t+\Delta,t)$ is the number of events occurring between $t+\Delta$ and $t$.

This number is evaluated by adopting the Feller prescription \cite{reference,fellerbook}. We assume that the system is prepared at time $t = 0$, we use  for $R(t)$ Eq. (\ref{feller})
and as a consequence the total number of events produced moving from $0$ to $\tau >0$, $m(\tau)$, is
\begin{equation}
m(\tau) = \int_{0}^{\tau} \frac{A}{t'^{1-\alpha}} dt' = \frac{A}{\alpha} \tau^{\alpha}. 
\end{equation}
The number $n(t+\Delta,t)$ is given by
\begin{equation} \label{thesameasthat}
n(t+\Delta,t) = \frac{A}{\alpha} \left[(t+\Delta)^{\alpha} - t^{\alpha}\right].
\end{equation}
We make the assumption $\Delta \ll t$. With this assumption we are allowed to use a Taylor series expansion and to neglect the  higher order terms, thereby  getting
\begin{equation} \label{betweentandt+Delta}
n(t+\Delta,t) \sim A t^{\alpha -1} \Delta.
\end{equation}
By plugging Eq. (\ref{betweentandt+Delta}) into Eq. (\ref{ensemble})
we get, after a straightforward time integration, 
\begin{equation} \label{barkai1}
\left<\overline{\delta^2}(\Delta)\right>= \frac{B \Delta^{2H}}{L(1 -\gamma) }  \left[L^{1-\gamma} - L_0^{1-\gamma}\right], 
\end{equation}
where 
\begin{equation} \label{definitionofgamma}
\gamma \equiv 2H(1-\alpha).
\end{equation}
Note that 
\begin{equation} \label{valueofB}
B = c A^{2H}.
\end{equation}
We introduce this factor so as to recover the key result of Eq. (\ref{ensemble}) when $H = 0.5$.

When $\gamma < 1$ and $L \gg L_0$ Eq. (\ref{barkai1}) yields
\begin{equation} \label{final0}
\left<\overline{\delta^2}(\Delta)\right> = \frac{B}{L^{2H(1-\alpha)}} \frac{1}{1 - 2H(1-\alpha)} \Delta^{2H}. 
\end{equation}
This equation is a generalization of the important result of Eq. (\ref{definition}). In fact, when $H = 0.5$ Eq. (\ref{final0}) becomes identical to Eq. (\ref{definition}), which is the main result of the theoretical proposal of Ref. \cite{reference}.
Another interesting property of Eq. (\ref{final0}) is that for $\alpha = 1$ the second moment $\overline{\delta^2}(\Delta)$ becomes independent of the length $L$ of the time series and identical to the second moment evaluated with the conventional ensemble average, thereby showing that FBM is ergodic.
If $\alpha < 1$, the cascade of Feller events makes the intensity of $\overline{\delta^2}(\Delta)$, namely the factor of $\delta^{2H}$ in Eq.  (\ref{final0}), depend on the time length of the sequence. This form of ergodicity breaking does not affect the FBM scaling.  

\begin{figure}[ht]

\includegraphics[width= 90mm]{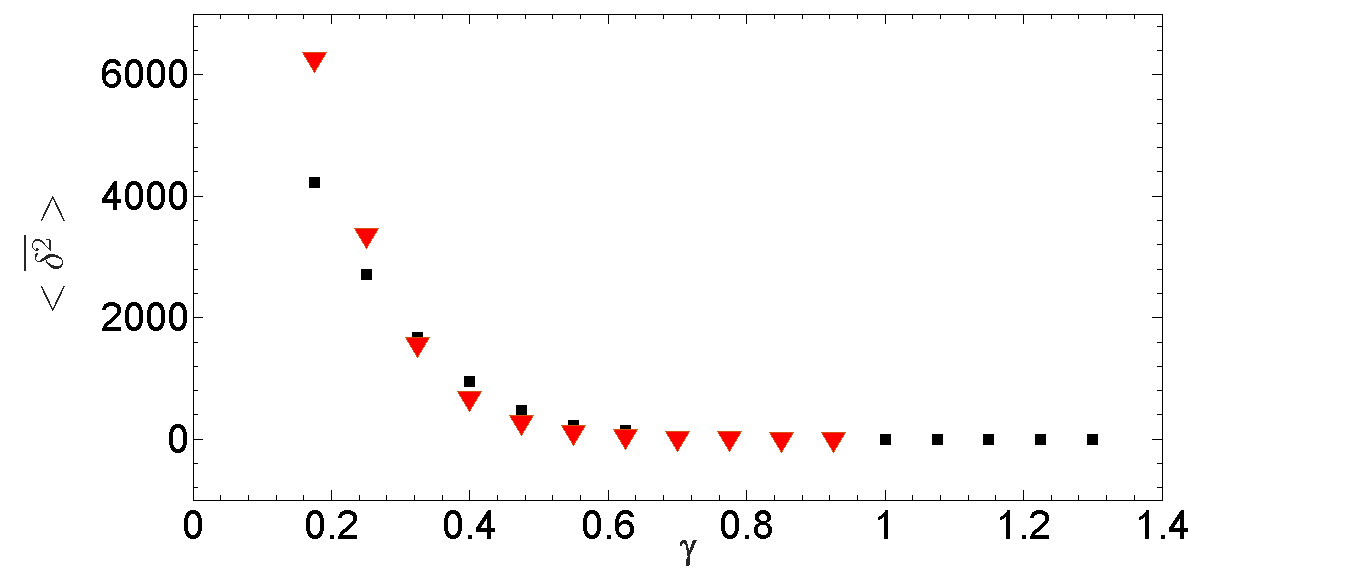}
 \caption{The red triangles denote $\left<\overline{\delta^2}(\Delta)\right>$ of Eq. (\ref{final0}) as a function of $\gamma$ given by Eq. (\ref{definitionofgamma}). The ensemble average is done on $10$ single trajectory realizations.  We set $H = 0.75$, $\Delta = 10^3$, $L = 10^6$. To define Eq. (\ref{final0}) we use  Eq. (\ref{definitionofdelta}), Eq. (\ref{definitionofc}), Eq. (\ref{valueofA}), Eq. (\ref{valueofB}) and Eq. (\ref{lambda}).  Eq. (\ref{definitionofdelta}) with $H = 0.75$ yields $\nu = 0.5$. For Eq. (\ref{definitionofc}) we use $\left<\xi^2\right> = 0.0657$, $T_B = 0.129$, yielding $c = 0.0627$.  Eq. (\ref{valueofA}) depends on Eq. (\ref{lambda}) with $\alpha$ changing with $\gamma$ and $T = 0.091$. The black squares show the result of numerical treatment and for $\gamma > 1$ they agree with the vanishing value predicted by Eq. (\ref{barkai4}).}
\label{rohisha}\end{figure}

Note also that, as mentioned in Section \ref{introduction}, to maintain the structure of Eq. (\ref{definition}) we have to assign to the slowness index $\nu$ of Eq. (\ref{correlation2}) a value larger than a critical value $\nu_c$ .  We are now in a position to define this critical value of $\nu$.  First of all we notice that 
 $\gamma > 1$ is possible for $\alpha < 0.5$. In fact the critical value of $H$, $H_c$,  at which $\gamma = 1$,  is given by
\begin{equation}
H_c = \frac{1}{2(1-\alpha)}.
\end{equation}
To realize the condition $\gamma > 1$, we need to make 
$H$ larger than $H_c$ and this is impossible to do with $\alpha = 0.5$, a value of $\alpha$ assigning to $H$ its  maximal value $H = 1$. Larger values of $\alpha$ would 
make $H$ exceed this maximal value. The adoption of $\alpha < 0.5$ generates $H_c  < 1$, and consequently, according to Eq. (\ref{definitionofgamma}), 
 $\gamma < 1$.  The  violation of the condition $\gamma < 1$ requires $H > H_c$ and a slowness parameter $\nu$ small enough. Using Eq. (\ref{definitionofdelta}) we are led to define $\nu_c$ as
\begin{equation} \label{definitionofnucritical}
\nu_c  = \frac{1 - 2 \alpha}{1 - \alpha}.
\end{equation} 
The condition $\gamma < 1$ corresponds to $\nu > \nu_c$. Adopting a correlation function with a slower decay, $\nu < \nu_c$, is equivalent to setting $\gamma > 1$, and this has the dramatic effect of making $\overline{\delta^2}(\Delta)$ vanish.
In fact, with $\gamma > 1$ Eq. (\ref{barkai1}) becomes
\begin{equation} \label{barkai2}
\left<\overline{\delta^2}(\Delta)\right> = \frac{B \Delta^{2H}}{L(\gamma -1) }  \left[\frac{1}{L_0^{\gamma-1}} - \frac{1}{L^{\gamma-1}}\right]. 
\end{equation}
It is convenient to write this expression in the form
\begin{equation} \label{barkai4}
\left<\overline{\delta^2}(\Delta)\right>=  \frac{L_0}{L}
\frac{B\Delta^{2H}}{L_0^{\gamma} (\gamma-1)} \left[1 - \left(\frac{L_0}{L}\right)^{\gamma-1}\right],\end{equation}
which shows that for $L_0/L \rightarrow 0$ the second moment $\overline{\delta^2}(\Delta)$ vanishes, yielding a perfect localization, in the same way as Eq. (\ref{definition}) does for $\alpha \rightarrow 0$, as the readers can easily realize from \cite{note} and from the property $\lim_{\alpha \rightarrow 0} \Gamma (\alpha) = \infty$.  In the case of ordinary CTRW this localization is a natural consequence of the fact that with $\alpha \rightarrow 0$ for the entire time interval $L$ no new event may occur in addition to the preparation event.  The same  effect emerges from the joint action of NPR events and FBM fluctuation, and this is the reason why we use the term localization to define it. However, the localization of this paper is not a consequence of lack of events, since, as shown in Section \ref{eventsoccur}, the ensemble average approach yields a second moment of the diffusion process increasing in time, although with a very small scaling. The localization effect of this paper, produced  by the joint action of NPR events and FBM fluctuation, is determined  by the FBM infinite memory transmitted to the NPR events. The rare occurrence of NPR events makes the slow decay of the correlation function of Eq. (\ref{correlation2}) become infinitely slow and the single trajectories become equivalent to deterministic and ballistic trajectories, with the effect of  annihilating $\overline{\delta^2}(\Delta)$. 
This is the reason why we refer to this effect as \emph{memory-induced localization}. 
The theoretical prediction of memory-induced localization is satisfactorily supported by the numerical  calculations of Fig. \ref{rohisha}.

It is important to point out that the theoretical prediction of this Section can be interpreted as being based on the assumption 
\begin{equation} \label{wrong}
\left <[x(t) - x(t+\Delta)]^2 \right> \propto \left<n(t+\Delta,t)\right>^{2H}.
\end{equation}
 As pointed out by the authors of Ref. \cite{sokolov3} the number of events 
between $t+\Delta$ and $t$ is not fixed  and can be evaluated by adopting the following prescription
\begin{equation} \label{rightprescription}
\left<(n(t+\Delta) - n(t))^{2H}\right>  = \int_{0}^{\Delta} d\tau \psi_{t}(\tau) n(t+\tau)^{2H}, 
\end{equation}
where $\psi_{t}(\tau)$ is the waiting time distribution density when we begin waiting at time $t$ far from the occurrence of the preparation event.  
Following the calculations done in Ref. \cite{sokolov3} we get
\begin{equation} \label{sokolovisright}
\left<\overline{\delta^2}(\Delta)\right> \propto \frac{1}{L^{1-\alpha}} \Delta^{1- \alpha + 2H \alpha}.  
\end{equation}
As we see, with the adoption of the prescription of Eq. (\ref{rightprescription}) the signature of the joint action of NPR events and FBM generating fluctuations is given by the power index of $\Delta$ which is $1- \alpha + 2H \alpha$ rather than $2H$. There is no memory-induced localization. 
  
\begin{figure}[h]

\includegraphics[width= 90mm]{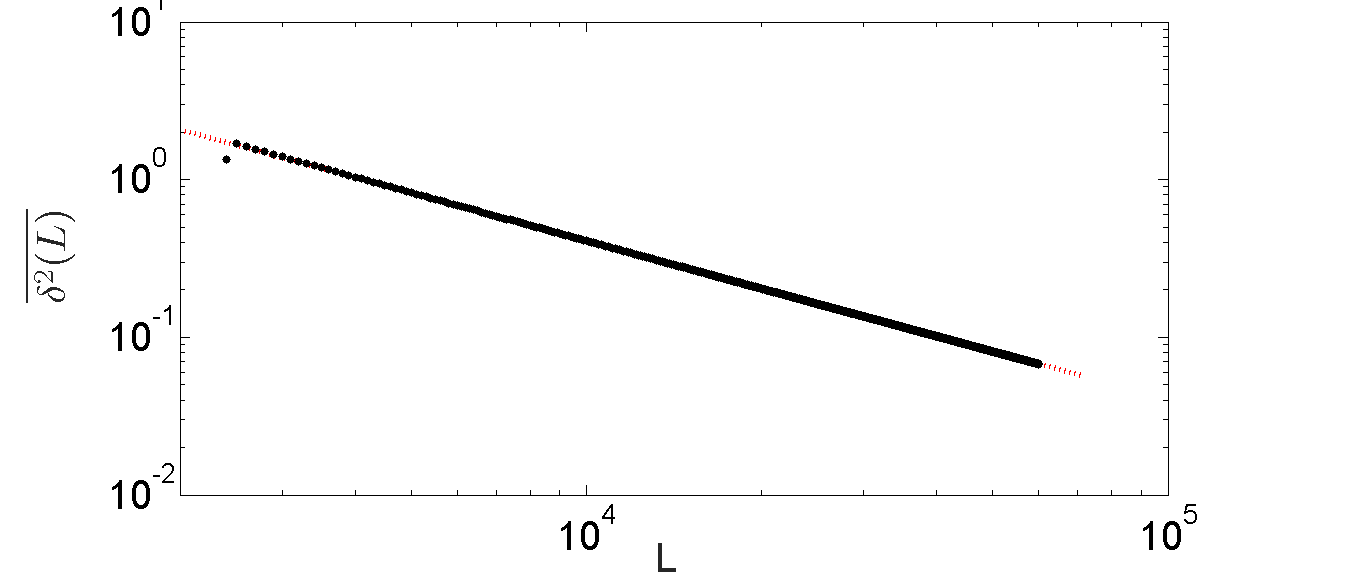}
 \caption{The single trajectory time average evaluation of the second moment of Eq. (\ref{ideal} ) as a function of $L$. $H = 0.75$, $\alpha = 0.35$, $\Delta = 100$. The black dots denote the numerical results and the dashed red line is the fitting yielding the slope $1.0057$. According to Eq. (\ref{final0}) the theoretical slope of the straight line is $2H(1-\alpha) = 0.975$. }
\label{ROHISHAisRIGHT}\end{figure}
\begin{figure}[ht]

\includegraphics[width= 90mm]{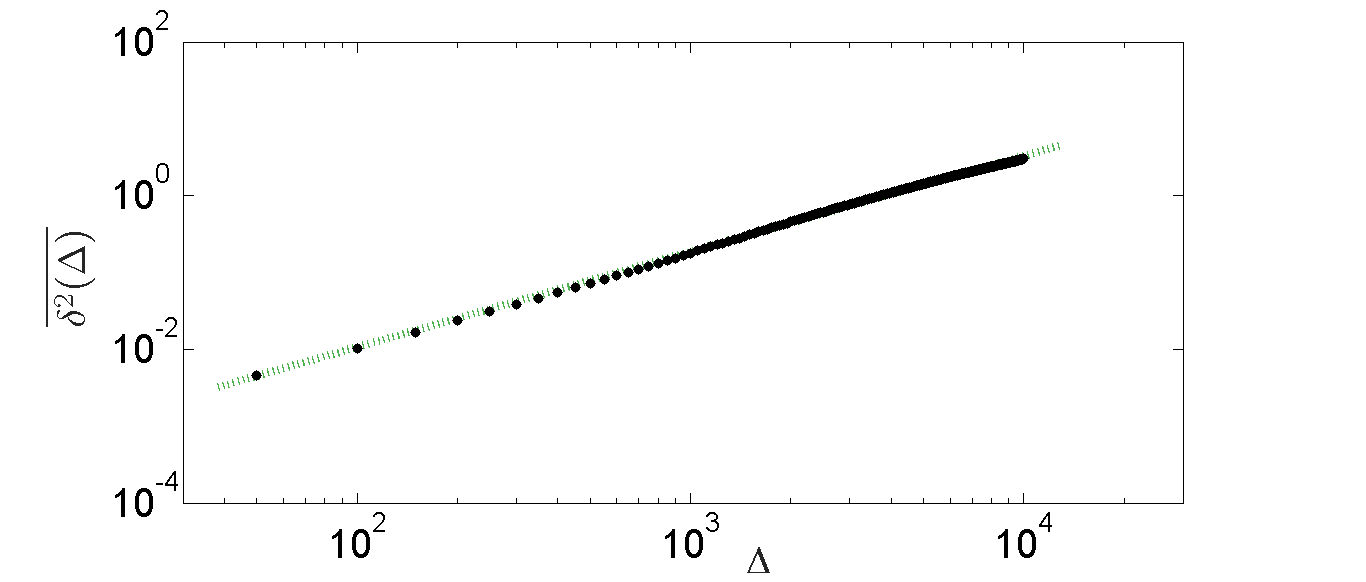}
 \caption{The single trajectory time average evaluation of the second moment of Eq. (\ref{ideal} ) as a function of $\Delta$. $H = 0.75$, $\alpha = 0.4$, $L = 10^6$. The black dots denote the numerical results and the dashed green line is the fitting yielding the slope $1.244$. According Eq. (\ref{sokolovisright}) the slope of the straight line is $1 - \alpha + 2 \alpha  H = 1.2$. Note $2H = 1.5$. }\label{ROHISHAisRIGHT2}
 \end{figure}

To support our arguments on the memory-induced localization, in spite of the striking difference with the rigorous result of Ref. \cite{sokolov3} 
we  make the numerical calculation of Fig. \ref{ROHISHAisRIGHT}. This figure refers to the case where $\gamma$ is very close to the border limit $\gamma = 1$ and we think that the agreement between numerical results and theoretical prediction is satisfactory. 
However, Fig. \ref{ROHISHAisRIGHT2} shows that the theoretical prediction of Ref. \cite{sokolov3} on the power index of $\Delta$ is correct. In the concluding remarks we shall come back to discuss the conflict between the predictions of this article and the earlier work of Ref. \cite{sokolov3}.

\subsection{Ensemble average} \label{eventsoccur}

In the operational time scale the diffusion process is described by
\begin{equation} \label{psycho}
p_{\tau_{\psi}}(x, \tau_{\psi}) =  \frac{1}{\sqrt{2 \pi k \tau_{\psi}^{2H} }} \exp\left(- \frac{x^2}{2 k \tau_{\psi}^{2H} }\right).
\end{equation}
Moving from the operational to the clock time we get

\begin{equation} \label{clocktime}
p_{C}(x,t) = \int_{0}^{\infty} d\tau_{\psi} \beta(t, \tau_{\psi}) \frac{1}{\sqrt{2 \pi k \tau_{\psi}^{2H} }} \exp\left(- \frac{x^2}{2 k \tau_{\psi}^{2H} }\right),
\end{equation}
where $\beta(t, \tau_{\psi})$ denotes the distribution density of time $t$ interpreted as the diffusion variable of the discrete time $n$ that in the asymptotic limit becomes 
the continuous variable $\tau_{\psi}$ \cite{ewa,russian}. 
The clock time $t$ is related to the operational time $\tau_{\psi}$ by the relation of Eq. (\ref{truescaling}),
which is a consequence of the fact that $t$ plays the role of diffusional variable. This yields
\begin{equation} \label{second2}
  \tau_{\psi} \propto t^{\alpha}.
\end{equation}
We can find the scaling $\eta$ generated by Eq. (\ref{clocktime}) with the following intuitive argument. The distribution $\beta(t, \tau_{\psi})$ sets the constraint $\tau_{\psi} = t^{\alpha}$. As a consequence 
the variable $x$, which is proportional to $\tau_{\psi}^{2H}$, turns out to be proportional to $t^{\eta}$, 
with
\begin{equation} \label{notrivial}
\eta = \alpha H.
\end{equation}

This result can be obtained in a more rigorous way by noticing that the scaling of Eq. (\ref{second2}) corresponds to 
\begin{eqnarray}  \label{finalscaling}
\beta(t,\tau_{\psi})=\frac{1}{t^{\alpha}}F\left(\frac{\tau_{\psi}}{t^{\alpha}}\right).
\end{eqnarray}
By plugging $\beta(t, \tau_{\psi})$ of Eq. (\ref{finalscaling}) into Eq. (\ref{clocktime}) and replacing the integration variable
$\tau_{\psi}$ with $y \equiv \tau_{\psi}/t^{\alpha}$, we obtain
\begin{equation} \label{clocktime2}
p_{C}(x,t) = \frac{1}{t^{\alpha H}}  g_{H}\left(\frac{x^2}{t^{2H\alpha}}\right)   ,
\end{equation}
where
\begin{equation}
g_{H}(z) \equiv \int_{0}^{\infty} dy F(y) \frac{1}{\sqrt{2\pi y^{2H}}} \exp\left[- \frac{z^2}{2ky^{2H}}\right],
\end{equation}
thereby yielding a more rigorous derivation of the scaling of Eq. (\ref{notrivial}) .

Note that thanks to the earlier work of Ref. \cite{central} the Mittag-Leffler function is connected to $\beta(t, \tau_{\psi})$  through 
\begin{equation}
E_{\alpha}\left(-t^{\alpha} s\ \right)  =  \int_{0}^{\infty} d\tau_{\psi} \beta(t, \tau_{\psi}) \exp(-\tau_{\psi} s) .
\end{equation}
for $s \rightarrow 0$. 
As a consequence the double Laplace transform of $\beta(t,\tau_{\psi})$, $\hat \beta(u,s)$, reads
\begin{equation} \label{totransform}
\hat \beta(u,s) = \frac{1}{u + u^{1-\alpha}s}.
\end{equation}
In the special case $\alpha = 0.5$ inverse-Laplace transforming Eq. (\ref{totransform}) gives
\begin{eqnarray}
 \beta(t,\tau_{\psi})=  \frac{\exp\left(- \frac{\tau_{\psi}^2}{4t}\right)}{\sqrt(\pi t)} .
 \end{eqnarray}
Using Eq. (\ref{clocktime}) we get

\begin{eqnarray}
p_C(x,t)=\int_{0}^{\infty}\frac{\exp\left(- \frac{\tau_{\psi}^2}{4t}\right)}{\sqrt{\pi t}} \frac{\exp\left[-\frac{x^2}{2 \tau_{\psi}^{2H}}\right]}{\sqrt{2\pi \tau_{\psi}^{2H}}}d\tau_{\psi}.
\end{eqnarray}

On the basis of earlier arguments it is straightforward to prove that this equation can be written under the form
\begin{equation}
p_C(x,t) = \frac{1}{t^{\alpha H}} F\left(\frac{x}{t^{\alpha H}}\right).
\end{equation}
For example, in the case $H=1$ we get
\begin{equation} \label{mauro}
p_C(x,t) = \frac{1}{\pi\sqrt{2t}} K_0\left(\frac{\mid x \mid}{\sqrt{2t}}\right),
\end{equation}
where $K_0(z)$ is the modified Bessel function of the second kind.
Note that Eq. (\ref{mauro}) refers to the case $\alpha = 0$ and $H = 1$, namely the case where $\gamma$ of Eq. (\ref{definitionofgamma}) is $\gamma = 2$. This is a condition where $\left<\overline{\delta^2}(\Delta)\right> = 0 $, according to the theoretical and numerical results of Section \ref{timeaverage}. In fact, as stressed in Section \ref{mainresult}, localization occurs when $\gamma > 1$. The localization of a single trajectory is a clear manifestation of ergodicity breaking, because the localization of a single trajectory is derived from the statistical analysis in time, departing from the ensemble average analysis yielding no localization. Ergodicity breaking can also be realized as ensemble and time analysis bringing two different scaling. We notice, in fact, that the scaling of  Eq. (\ref{mauro}), ensemble analysis,  is $\eta = 1/2$, in accordance with Eq.(\ref{notrivial}), whereas the localization $\left<\overline{\delta^2}(\Delta)\right> = 0 $, observed doing time analysis, suggests that the single trajectories have the scaling $\eta = 0$.

\subsection{Concluding remarks on this first form of joint action}
We see that this form of joint action generates ergodicity breaking. In fact,  the ensemble average analysis yields Eq. (\ref{final0}) and consequently the scaling of Eq. (\ref{notrivial}),  $\eta = \alpha H$.  The time average analysis, on the contrary, generates a scaling dependent on whether the prescription of Eq. (\ref{wrong}) or the prescription of Eq. (\ref{rightprescription}) is adopted. 
In the former case
\begin{equation} \label{doubts}
\eta = H
\end{equation}
and in the latter
\begin{equation} \label{manyrealizations}
\eta = H\alpha + \frac{1 - \alpha}{2},
\end{equation}
thereby yielding ergodicity breaking regardless of what 
prescription is adopted. 
The three scalings are identical only in the absence of the Feller cascade, $\alpha = 1$, a property  reminiscent of the results of Refs. \cite{sokolov1,reference}. 

From a theoretical point of view, averaging on infinitely many single trajectory realization casts some doubts on the prescription of Eq. (\ref{wrong}) and especially on the scaling of Eq. (\ref{doubts}). However, the observation of single trajectories, with the adoption of time averages, 
shows that the dependence on $L$ of Eq. (\ref{final0}) is 
a correct prediction, thereby supporting our conclusion that $\gamma > 1$ generates localization, whereas the ensemble average generates a sub-diffusion process more pronounced than that of CTRW, with nevertheless the non-vanishing scaling $\eta = \alpha H$.

  \begin{figure}[ht]

\includegraphics[width= 90mm]{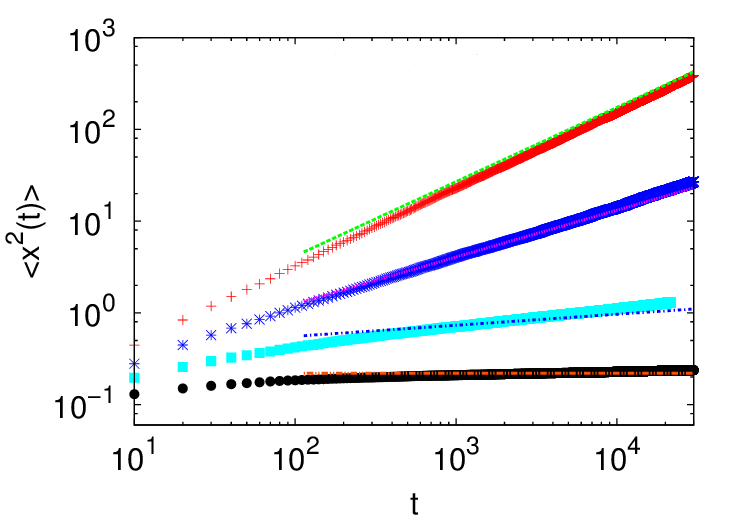}
 \caption{ The ensemble average of the second moment $\left<x^2(t)\right>$ for different values of $H$ and $\alpha$. All the curves were generated from the average of $10^4$ realizations. From the top to the bottom: $H = 0.45, \alpha = 0.9$; $H = 0.45, \alpha = 0.6$; $H = 0.10, \alpha = 0.9$; $H = 0.10, \alpha = 0.5$. All the straight lines but the bottom one are guidelines indicating the scaling $2 \eta = (2H-1 + \alpha)$. Note that  for the bottom curve, corresponding  to $2 \eta = -0.2$, the numerical result fits  the guideline straight curve with vanishing slope. }
\label{mekla2time2}.
\end{figure}

  \begin{figure}[ht]

\includegraphics[width= 90mm]{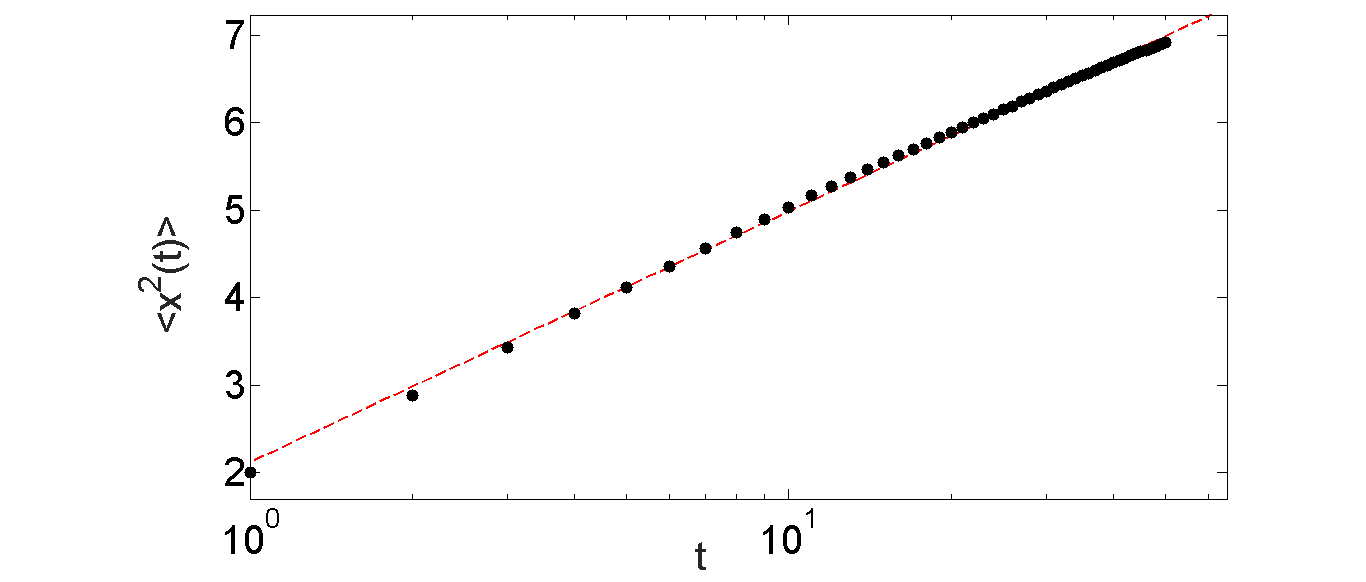}
 \caption{ Numerical evidence that $\eta < 0 $ makes $x$ proportional to $ln t$. $H = 0.1, \alpha = 0.6$. The black dots are the numerical results and the red dashed line is a linear-log fitting with the slope $1.23$.  Note that with these values the scaling of Eq. (\ref{crucialscaling}) becomes negative,  $\eta = -0.1. $}
\label{mirza}
\end{figure}

\section{Joint action of NPR events and SBM fluctuations} \label{secondform}

The second form of joint action is realized by adopting SBM rather than FBM, and by assuming that SBM is activated
in the clock time. The runner, however, spends a large part of her time sleeping. She jumps only when she is awake. 

The ensemble average is established through the formula
\begin{equation} \label{letususeagain2}
\left<x^2(t)\right> = \int_{L_0}^{t} dt_1 \int_{L_0}^{t} dt_2 <\xi^{*}(t_1)\xi^{*}(t_2)>,
\end{equation}
where $\xi^{*}(t)$  denotes the fluctuation perceived by the runner. This fluctuation is different from the vanishing value only when
the runner is awake. The sleeping condition is equivalent to setting $\xi^{*}(t)=0$. In this case the joint action of the Feller cascade and the SBM  diffusion time dependence
yields
\begin{equation} \label{trivial}
\left<\xi^{(*)}(t) \xi^{(*)}(t+ \tau)\right>= \frac{2kA}{t^{1-\alpha}} t^{2H-1} \delta(\tau),
\end{equation}
with the constant $k$ being determined by the prescription of Eq. (\ref{prescriptionofsokolov}) and the constant $A$ by the prescription of Eq. (\ref{valueofA}). 
In this section we focus only on the scaling at the level of ensemble statistics and at the level of time statistics. Therefore, for simplicity's sake we set $2kA =1$.  
By plugging Eq. (\ref{trivial}) with $2kA = 1$ into Eq. (\ref{letususeagain2}), we obtain
\begin{equation} \label{barkai3}
\left<x^2(t)\right> = \frac{1}{2H-1 + \alpha} \left[t^{2H-1+\alpha} - L_0^{2H-1+\alpha}\right].
\end{equation}

Notice that 
\begin{equation}
2H-1 + \alpha = 2 \eta,
\end{equation}
where 
\begin{equation} \label{crucialscaling}
\eta = \frac{2H - 1 + \alpha}{2}.
\end{equation}
This is the scaling generated by the generalized diffusion equation proposed by the authors of Ref. \cite{bologna}. 
When $\eta > 0$ and $t \gg L_0$, Eq. (\ref{barkai3}) becomes
\begin{equation} \label{elvisbarkai1}
\left<x^2(t)\right> = \frac{1}{2\eta} t^{2\eta}.
\end{equation}

It is worth noting that Ref. \cite{bologna} does not discuss  the dynamics of this process when 
$\eta < 0$.  
As a result of the theoretical approach illustrated in this paper we  reach the compelling conclusion that this negative scaling leads to an exact localization. In fact, 
when $\eta < 0$, Eq. (\ref{barkai3}) yields
\begin{equation} \label{elvisbarkai2}
\left<x^2(t)\right> = \frac{1}{2|\eta|} \left[\frac{1}{L_0^{2|\eta|}} - \frac{1}{t^{2|\eta|}}\right].
\end{equation} 
This prediction is supported by the numerical results of Fig. \ref{mekla2time2}. Notice that both Eq. (\ref{elvisbarkai1}) and Eq. (\ref{elvisbarkai2}) yield for $\eta \rightarrow 0$ the logarithmic scaling $x^2 \sim \ln t$. Fig. \ref{mirza} supports this prediction  in the case $\eta < 0$.

What about the analysis of single trajectories in this case?
To do the time average analysis we must adopt 

 \begin{equation} \label{newversion}
 \begin{split}
\left<\overline{\delta^2}(\Delta)\right> = \frac{1}{L} \int_{L_0}^{L} \left<\left[x(t) - x(t+\Delta)\right]^2 \right>dt \\
= \frac{1}{L} \int_{L_0}^{L} dt C(\Delta, t), 
\end{split}
\end{equation} 
where the function $C(\Delta,t)$ reads
\begin{equation}
 \label{correlation333}
C(\Delta, t) \equiv  \int_t^{t+\Delta} dt_1  \int_t^{t+\Delta} dt_2 \left<\xi^{(*)}(t_1)\xi^{(*)}(t_2)\right> ,
\end{equation} 
with the correlation function $\left<\xi^{(*)}(t_1)\xi^{(*)}(t_2)\right>$ given by 
\begin{equation} \label{lastcorrelation}
\left<\xi^{(*)}(t_1)\xi^{(*)}(t_2)\right>  = \frac{1}{t_1^{2-\mu}} t_1^{2H-1} \delta(|t_1-t_2|). 
\end{equation}
By plugging  Eq. (\ref{lastcorrelation}) into the right hand side of Eq. (\ref{correlation333}) we get
\begin{equation}
C(\Delta,t) = \frac{1}{2\eta} \left[(t+\Delta)^{2 \eta} - t^{2 \eta}\right].
\end{equation}
Doing the usual Taylor series expansion with the assumption
$\Delta < t$ and neglecting the terms of higher order in $\Delta/t$, we obtain
\begin{equation} \label{final}
C(\Delta,t) = \Delta t^{2\eta-1}.
\end{equation}
Plugging $C(\Delta,t)$ of Eq. (\ref{final})  into the right hand side of 
Eq. (\ref{newversion}) and integrating over $t$ we get
\begin{equation} \label{mirzaisright}
\left<\overline{\delta^2}(\Delta)\right> = \frac{\Delta}{2 \eta L}  \left[L^{2\eta} - L_0^{2\eta}\right].
\end{equation}
When $\eta > 0$ and $L \gg L_0$, we get
\begin{equation} \label{torewrite}
\left<\overline{\delta^2}(\Delta)\right> = \frac{1}{L^{1-2\eta}}\frac{\Delta}{2 \eta}.
\end{equation}
When $\eta < 0$ and $L \gg L_0$ Eq. (\ref{mirzaisright}) becomes
\begin{equation} \label{perfect}
\left<\overline{\delta^2}(\Delta)\right> = \Delta \left(\frac{L_0}{L}\right)^{2|\eta|}\left(1 - \left(\frac{L_0}{L}\right)^{2|\eta|}\right),
\end{equation}
yielding a vanishing second moment for $L_0/L \rightarrow 0$.

\section{Concluding Remarks} \label{concludingremarks}

This article affords a solution to the question raised by
the authors of Ref. \cite{bologna} on the interpretation of the scaling generated by the diffusion equation\begin{equation} \label{eta}
\frac{\partial^{\alpha}}{\partial t^{\alpha}} p(x,t)  = D(t) \frac{\partial^{2}}{\partial x^{2}} p(x,t), 
\end{equation}
where
the fractional derivative rests on the Caputo prescription \cite{caputo} 
and, according to Eq.(\ref{timedependentdiffusioncoefficient}),
\begin{equation} \label{eta1}
D(t) = c t^{2H-1}.
\end{equation}
The authors of \cite{bologna}  proved that this diffusion equation generates the scaling of Eq. (\ref{crucialscaling}) but they were unable to establish the nature of the individual stochastic trajectories.
An important result of this article is the discovery that the single trajectories corresponding to this scaling are those illustrated in Section \ref{secondform}. The research that we did to answer this question led us to establish that the condition $\eta \rightarrow 0$ yields the logarithmic scaling $x^2  \sim \ln t$, regardless of whether the vanishing value of $\eta$ is realized moving from $\eta >0$ or from $\eta < 0$. This process is characterized by ergodicity breaking and it is remarkable that the condition $\eta < 0$ corresponds to a perfect localization, as proved by Eq. (\ref{perfect}).

Much more surprising is the result that we obtain thanks to the joint action of NPR and FBM, as illustrated in Fig. 1. In this case the localization induced by ergodicity breaking is realized with $H > 0.5$ corresponding to super-diffusion. In the whole region of the memory-induced localization $\gamma >1$, the ensemble scaling $\eta = \alpha H$ applies and $\alpha < 0.5$ makes $\eta < 0.5$ even in the condition of maximal FBM super-diffusion, $H = 1$.  Thus, the whole region $\gamma > 1$ corresponds to ensemble sub-diffusion.  However, the memory-induced localization of this article is not a  trivial consequence of analyzing 
the time average of single trajectories when we have ensemble sub-diffusion.  In fact, the conditions $\alpha = 0.6$, for instance, with $H = 1$ yields $\eta = 0.6$, namely, super-diffusion,  and with $H = 0.75$  yields $\eta = 0.45$, namely, sub-diffusion.  Yet, both conditions are characterized by $\gamma < 1$, $\gamma = 0.8$, the former, and $\gamma =0.6$, the latter, and consequently,  in neither conditions memory-induced localization is produced.  In conclusion, we may have ensemble sub-diffusion and no localization  with time averages on the single trajectories.

The adoption of the prescription of Eq. (\ref{rightprescription}) adopted by the authors of Ref. \cite{sokolov3} casts some doubts on this effect, insofar as it yields the theoretical prediction of Eq. (\ref{sokolovisright}), with no localization effect. 
However, we note that this theoretical prediction
is based on the assumption  that the ensemble 
average commutes with the time integration, making 
Eq. (\ref{real}) turn into Eq. (\ref{ensemble}). This property is correct if the ensemble average of the time averages is done on infinitely many realizations. The 
numerical results of Fig. (\ref{rohisha}) are obtained making an ensemble average over ten single-trajectory realizations and the results of Fig. (\ref{ROHISHAisRIGHT}) and of Fig. (\ref{ROHISHAisRIGHT2}) are based on single trajectories. They show that the scaling prediction of Eq. (\ref{doubts}) is incorrect and that probably averaging over a suitably large number of realizations would lead 
to the scaling of Eq. (\ref{manyrealizations}). 
We believe that the correct theoretical prediction of Eq. (\ref{sokolovisright}) corresponds to the superposition of infinitely many single realizations, each of which is has an almost fixed non-vanishing value $\overline{\delta^{2}}$, namely, a localized single trajectory.

In conclusion,  the memory-induced localization is a single-trajectory property and 
we are convinced that this result may lead to a new vision of memory and renewal processes.  To shape this vision it may be convenient to consider FBM with the second moment 
increasing as $t^{2H}$ as an indication of randomness quite distinct from that signaled by the occurrence of renewal events. The correlation function of Eq. (\ref{correlation2}) is compatible with the derivation from 
the Hamilton formalism and consequently with the Laplace determinism \cite{biology,annunziato}. This suggests that the entropy increase in this case is due  to the lack of information generated by a contraction over the irrelevant variables that, in spite of being ignored, are responsible  for the slow decay of this correlation function. The second moment increasing as $t^{\alpha}$, on the contrary, has a completely different meaning. In the operational time,
in fact, any time step is characterized by the action of a random event, thereby forcing the second moment to increase  linearly with $\tau_{\psi}$. As a matter of fact, as we have seen, $t^{\alpha} = \tau_{\psi}$, in line with the extended definition of Lyapunov coefficient adopted by Korabel and Barkai to adapt it to the occurrence of rare events of the $\alpha < 1$ condition \cite{korabel}. In this case,
when we depart from the singularity condition $H = 0.5$, where $\xi$ is completely uncorrelated, and we move towards $H = 1$ with the decay of the correlation function of Eq. (\ref{correlation2}) becoming slower and slower, the correlated fluctuation $\xi$ has the effect of completely quenching 
the action of NPR randomness at the critical value $H_c = 1/{[2(1-\alpha)]}$, corresponding to the critical slowness of Eq. (\ref{definitionofnucritical}). Although we have been  referring ourselves to this phenomenon as memory-induced localization, imagining the FBM memory as the cause of the effect, this may be probably equivalent  to interpret the rare events of the condition $\alpha < 1$ as the cause of the effect, being a physical condition that makes the decay of correlation function of Eq. (\ref{correlation2}) infinitely slow.  

The settlement of these problems require further studies and a promising research direction is suggested by the recent work of Marzen and Crutchfield \cite{marzen} where infinite memory and renewal events are discussed using the mutual information between past and future and the amount of information from the past required to exactly predict the future.

 {\bf{ACKNOWLEDGMENT}} 
The authors thank Welch for financial support through Grant No. B-1577.


.


\begin{references}
\bibitem{note}  We note that the complete expression for Eq. (\ref{definition}) is $ \overline{\delta^2}(\Delta) = \left\{ \left<\xi^2\right>\Delta \right\}/\left\{\alpha T^{\alpha} \Gamma(\alpha) \Gamma(1-\alpha)L^{1-\alpha}\right\}$.
\bibitem{sokolov4} The very interesting work of Ref. \cite{sokolov} refers to the case where the diffusion process in the operational time scale occurs in a fractal environment where the second moment
$\left<x^2(t)\right> \propto t^\beta$. However, the theoretical results of this paper can be fruitfully adopted in the present article by replacing $\beta$ with $2H$, noticing that the authors of Ref. \cite{sokolov3} discussed the case where the diffusion process in the operational time is sub diffusional, $\beta < 1$ and consequently $H < 0.5$.
\end{references}
\end{document}